\input harvmac
\input tables

%
%
%

\let\includefigures=\iftrue
\let\useblackboard=\iftrue
\newfam\black

\def\Z{\relax\ifmmode\mathchoice
{\hbox{\cmss Z\kern-.4em Z}}{\hbox{\cmss Z\kern-.4em Z}}
{\lower.9pt\hbox{\cmsss Z\kern-.4em Z}}
{\lower1.2pt\hbox{\cmsss Z\kern-.4em Z}}\else{\cmss Z\kern-.4em Z}\fi}

\noblackbox

\def\IZ{\relax\ifmmode\mathchoice
{\hbox{\cmss Z\kern-.4em Z}}{\hbox{\cmss Z\kern-.4em Z}}
{\lower.9pt\hbox{\cmsss Z\kern-.4em Z}}
{\lower1.2pt\hbox{\cmsss Z\kern-.4em Z}}\else{\cmss Z\kern-.4em Z}\fi}
\def\IB{\relax{\rm I\kern-.18em B}}
\def\inbar{\vrule height1.5ex width.4pt depth0pt}
\def\IC{\relax\,\hbox{$\inbar\kern-.3em{\rm C}$}}
\def\ID{\relax{\rm I\kern-.18em D}}
\def\IE{\relax{\rm I\kern-.18em E}}
\def\IF{\relax{\rm I\kern-.18em F}}
\def\IG{\relax\hbox{$\inbar\kern-.3em{\rm G}$}}
\def\IGa{\relax\hbox{${\rm I}\kern-.18em\Gamma$}}
\def\IH{\relax{\rm I\kern-.18em H}}
\def\II{\relax{\rm I\kern-.18em I}}
\def\IN{\relax{\rm I\kern-.18em N}}
\def\IM{\relax{\rm I\kern-.18em M}}
\def\IK{\relax{\rm I\kern-.18em K}}
\def\IP{\relax{\rm I\kern-.18em P}}

\def\IR{\relax{\rm I\kern-.18em R}}
\def\I1{\relax{\rm 1\kern-.38em 1}}
\def\D{\Delta}

\font\cmss=cmss10
\font\cmsss=cmss10 at 7pt

\def\p{\partial}

\def\ie{{i.e$\,$}}
\def\eg{e.g.$\,$}
\def\frac#1#2{{#1 \over #2}}
\def\a{\alpha}

\def\eps{\epsilon}

\def\l{\lambda}
\def\r{\rho}
\def\s{\sigma}
\def\t{\theta}

\def\M{{\cal M}}
\def\N{{\cal N}}

\def\O{{\cal O}}

\def\subsubsec{\subsec}

\lref\WittenXJ{
 E.~Witten,
 ``Topological Sigma Models,''
 Commun.\ Math.\ Phys.\  {\bf 118}, 411 (1988).
}

\lref\LercheUY{
 W.~Lerche, C.~Vafa and N.~P.~Warner,
 ``Chiral Rings In N=2 Superconformal Theories,''
 Nucl.\ Phys.\ B {\bf 324}, 427 (1989).
}

\lref\WittenIG{
 E.~Witten,
 ``On The Structure Of The Topological Phase Of Two-Dimensional Gravity,''
 Nucl.\ Phys.\ B {\bf 340}, 281 (1990).
}

\lref\AdamsZY{
 A.~Adams, A.~Basu and S.~Sethi,
 ``(0,2) duality,''
 Adv.\ Theor.\ Math.\ Phys.\  {\bf 7}, 865 (2004)
 [arXiv:hep-th/0309226].
}

\lref\BlumenhagenVU{
 R.~Blumenhagen, R.~Schimmrigk and A.~Wisskirchen,
 ``(0,2) mirror symmetry,''
 Nucl.\ Phys.\ B {\bf 486}, 598 (1997)
 [arxiv:hep-th/9609167].
}

\lref\BlumenhagenTV{
 R.~Blumenhagen and S.~Sethi,
 ``On orbifolds of (0,2) models,''
 Nucl.\ Phys.\ B {\bf 491}, 263 (1997)
 [arXiv:hep-th/9611172].
}

\lref\SharpeWA{
 E.~R.~Sharpe,
 ``(0,2) mirror symmetry,''
 arXiv:hep-th/9804066.
}

\lref\WittenZZ{
 E.~Witten,
 ``Mirror manifolds and topological field theory,''
 arXiv:hep-th/9112056.
}

\lref\MorrisonFR{
 D.~R.~Morrison and M.~Ronen Plesser,
 ``Summing the instantons: Quantum cohomology and mirror symmetry in toric varieties,''
 Nucl.\ Phys.\ B {\bf 440}, 279 (1995)
 [arXiv:hep-th/9412236].
}

\lref\KapustinPT{
 A.~Kapustin,
 ``Chiral de Rham complex and the half-twisted sigma-model,''
 arXiv:hep-th/0504074.
}

\lref\WittenPX{
 E.~Witten,
 ``Two-Dimensional Models With (0,2) Supersymmetry: Perturbative Aspects,''
 arXiv:hep-th/0504078.
}

\lref\agm{
P.~S.~Aspinwall, B.~R.~Greene and D.~R.~Morrison,
``Calabi-Yau moduli space, mirror manifolds and spacetime topology  change in string theory,''
Nucl.\ Phys.\ B {\bf 416}, 414 (1994)
hep-th/9309097.
}
\lref\AspinwallCF{
P.~S.~Aspinwall and B.~R.~Greene,
``On the geometric interpretation of N=2 superconformal theories,''
Nucl.\ Phys.\ B {\bf 437}, 205 (1995)
[arXiv:hep-th/9409110].
}
\lref\AspinwallAY{
P.~S.~Aspinwall,
``The Moduli space of N=2 superconformal field theories,''
arXiv:hep-th/9412115.
}
\lref\VafaUZ{
 C.~Vafa,
 ``Topological mirrors and quantum rings,''
 arXiv:hep-th/9111017.
}
\lref\IqbalDS{
 A.~Iqbal, N.~Nekrasov, A.~Okounkov and C.~Vafa,
 ``Quantum foam and topological strings,''
 arXiv:hep-th/0312022.
}
\lref\AspinwallZD{
P.~S.~Aspinwall, B.~R.~Greene and D.~R.~Morrison,
``Space-Time Topology Change And Stringy Geometry,''
J.\ Math.\ Phys.\  {\bf 35}, 5321 (1994).
}
\lref\AspinwallUJ{
P.~S.~Aspinwall and D.~R.~Morrison,
``Chiral rings do not suffice: N=(2,2) theories with nonzero fundamental group,''
Phys.\ Lett.\ B {\bf 334}, 79 (1994)
[arXiv:hep-th/9406032].
}

\lref\KatzNN{
 S.~Katz and E.~Sharpe,
 ``Notes on certain (0,2) correlation functions,''
 arXiv:hep-th/0406226.
}

\lref\SharpeFD{
 E.~Sharpe,
 ``Notes on correlation functions in (0,2) theories,''
 arXiv:hep-th/0502064.
}


\lref\KachruPG{
 S.~Kachru and E.~Witten,
 ``Computing the complete massless spectrum of a Landau-Ginzburg orbifold,''
 Nucl.\ Phys.\ B {\bf 407}, 637 (1993)
 [arXiv:hep-th/9307038].
}

\lref\DistlerMK{
 J.~Distler and S.~Kachru,
 ``(0,2) Landau-Ginzburg theory,''
 Nucl.\ Phys.\ B {\bf 413}, 213 (1994)
 [arXiv:hep-th/9309110].
}

\lref\BlumenhagenEW{
 R.~Blumenhagen, R.~Schimmrigk and A.~Wisskirchen,
 ``The (0,2) Exactly Solvable Structure of Chiral Rings, Landau-Ginzburg Theories, and Calabi-Yau Manifolds,''
 Nucl.\ Phys.\ B {\bf 461}, 460 (1996)
 [arXiv:hep-th/9510055].
}

\lref\KachruWM{
 S.~Kachru and C.~Vafa,
 ``Exact results for N=2 compactifications of heterotic strings,''
 Nucl.\ Phys.\ B {\bf 450}, 69 (1995)
 [arXiv:hep-th/9505105].
}


\lref\DistlerEE{
 J.~Distler and B.~R.~Greene,
 ``Aspects Of (2,0) String Compactifications,''
 Nucl.\ Phys.\ B {\bf 304}, 1 (1988).
}
\lref\DistlerMI{
 J.~Distler,
 ``Notes on (0,2) superconformal field theories,''
 arXiv:hep-th/9502012.
}

\lref\DineZY{
 M.~Dine, N.~Seiberg, X.~G.~Wen and E.~Witten,
 ``Nonperturbative Effects On The String World Sheet,''
 Nucl.\ Phys.\ B {\bf 278}, 769 (1986).
}
\lref\DineBQ{
 M.~Dine, N.~Seiberg, X.~G.~Wen and E.~Witten,
 ``Nonperturbative Effects On The String World Sheet. 2,''
 Nucl.\ Phys.\ B {\bf 289}, 319 (1987).
}
\lref\SilversteinRE{
 E.~Silverstein and E.~Witten,
 ``Criteria for conformal invariance of (0,2) models,''
 Nucl.\ Phys.\ B {\bf 444}, 161 (1995)
 [arXiv:hep-th/9503212].
}
\lref\BasuBQ{
 A.~Basu and S.~Sethi,
 ``World-sheet stability of (0,2) linear sigma models,''
 Phys.\ Rev.\ D {\bf 68}, 025003 (2003)
 [arXiv:hep-th/0303066].
}
\lref\BeasleyFX{
 C.~Beasley and E.~Witten,
 ``Residues and world-sheet instantons,''
 JHEP {\bf 0310}, 065 (2003)
 [arXiv:hep-th/0304115].
}

\lref\WittenYC{
 E.~Witten,
 ``Phases of N = 2 theories in two dimensions,''
 Nucl.\ Phys.\ B {\bf 403}, 159 (1993)
 [arxiv:hep-th/9301042].
}

\lref\HoriKT{
 K.~Hori and C.~Vafa,
 ``Mirror symmetry,''
 arXiv:hep-th/0002222.
}

\lref\MorrisonYH{
 D.~R.~Morrison and M.~R.~Plesser,
 ``Towards mirror symmetry as duality for two dimensional abelian gauge theories,''
 Nucl.\ Phys.\ Proc.\ Suppl.\  {\bf 46}, 177 (1996)
 [arXiv:hep-th/9508107].
}

\lref\SheldonWIP{
 S.~Katz et al, To Appear
}

%
\def\QR{{\overline Q}}
\def\QRB{{\overline Q}^\dagger}
\def\JR{{\overline J}}
\def\TR{{\overline T}}
\def\GR{{\overline G}}
\def\GRB{{\overline G}^\dagger}
\def\LR{{\overline L}}
\def\pzb{{\partial_{\overline z}}}
%
\def\QL{Q}
\def\QLB{Q^\dagger}
\def\JL{J}
\def\TL{T}

\def\LL{L}

%
%
%
%
\def\ib{{\overline \imath}}

\def\DB{{\overline \Delta}}
\def\V{{\cal V}}

\def\hb{\overline{h}}
\def\qb{\overline{q}}
\def\zb{{\overline z}}
\def\rk{{\rm rk}}
\def\sb{{\overline s}}

\Title{\vbox{\baselineskip12pt
\hbox{hep-th/0506263}\hbox{HUTP-05/A0026}\hbox{UTTG-06-05}}}{\vbox{
\centerline{Topological Heterotic Rings}
\medskip
}}
\centerline{Allan Adams$^1$, Jacques Distler$^2$ and Morten Ernebjerg$^1$}
\bigskip
\bigskip
\bigskip
\centerline{\it $^1$Jefferson Physical Laboratory, Harvard University, Cambridge, MA 02138}
\smallskip
\centerline{\it $^2$Department of Physics, University of Texas, Austin, TX 78712}

\bigskip
\bigskip
\bigskip
\bigskip
\bigskip
\bigskip

\noindent

We prove the existence of  topological rings in (0,2) theories
containing non-anomalous left-moving U(1) currents by which
they may be twisted.  While the twisted models
are not topological, their ground operators
form a ring under non-singular OPE which reduces to the (a,c) or
(c,c) ring at (2,2) points and to a classical sheaf cohomology ring at large radius,
defining a quantum sheaf cohomology away from these special loci.
In the special case of Calabi-Yau compactifications,
these rings are shown to exist globally
on the moduli space
if the rank of the holomorphic bundle is less than eight.

\Date{June 2005}

\newsec{Introduction}

$\N$=(2,2) sigma models provide a beautiful and well-padded
playground in which to study the quantum geometry of
spacetime\foot{For an absurdly partial list of reference see
\WittenXJ\agm\AspinwallCF\AspinwallAY\VafaUZ\IqbalDS\ and references therein}.
Among their most remarkable toys are their chiral rings, which
reduce to the classical cohomology rings of the target in the large
volume limit, defining a quantum cohomology ring at finite radius
where worldsheet instanton correct the classical result \LercheUY\WittenIG.

Since generic (2,2) models may be smoothly deformed into (0,2)
models\foot{For background on (0,2) models, see \eg
\WittenYC\DistlerEE\DistlerMI}, it is natural to wonder if there
is a (0,2) generalization of the quantum cohomology ring which
reduces to the finite-dimensional (a,c) ring on the (2,2) locus.
At first glance
this seems unlikely, since the space of chiral operators is
infinite dimensional in the absence of left-moving
supersymmetry.
Moreover, while (2,2) supersymmetry ensures the
non-singularity of the OPEs of (a,c) chiral operators (from which
topological invariance of their correlators follows), general
right-chiral operators in (0,2) models have singular (and thus
metric-dependent) OPEs.  Finally, worldsheet instanton effects can be much more
dangerous and uncontrolled than in the more well-understood (2,2) models - indeed,
until recently \SilversteinRE\BasuBQ\BeasleyFX, it was widely believed that most (0,2) models
were destabilized by instantons and could not be defined non-perturbatively \DineZY\DineBQ.
On the other hand, finite-dimensional
apparently topological rings have been identified in several (0,2) theories, both
by mirror symmetry \AdamsZY\ and through explicit construction in certain
exactly solved (0,2) models \BlumenhagenEW.
The problem is thus not {\it if} they exist, but {\it when}.

Classical geometry provides a hint in the case of (0,2) NLSMs on holomorphic vector
bundles over K\"ahler targets, $\V\to X$, where the right-moving supercharge maps to
the Dolbeault operator on $X$ twisted in the bundle $\V$ to which the left-moving
fermions couple.  In these models, while the cohomology of the right-moving
supercharge is in general infinite dimensional (and is related to the elliptic
cohomology of $X$), $H^*(X,\wedge^*\V)$ forms a finite dimensional sub-algebra.
When $\V$ is a smooth deformation of $T_X$, this is a smooth deformation of the
de Rham cohomology ring of $X$, and the usual trace, given by integration over
$X$, is maintained.  For more general $\V$, integration on $X$ provides a natural trace
if $\wedge^{top}\V=K_X^*$; the existence of a trace makes our ring a Frobenius
ring.  Suggestively, $\wedge^{top}\V=K_X^*$ implies the preservation of a left-moving
U(1) current algebra on the worldsheet; if the (0,2) theory is a deformation of a (2,2)
theory by an element of $H^1(X,End(T_X))$, this U(1) is the unbroken left-moving
R-symmetry.  It is thus natural to suppose that the ring of operators which computes
the de Rham cohomology in the classical limit of a (2,2) model persists as a
ring away from (2,2) loci.

In this paper we prove the existence of finite-dimensional topological rings in (0,2)
theories containing conserved left-moving $U(1)$ currents. While the A and B twists
are only quasi-topological away from (2,2) loci, their ground rings are
fully topological on open sets in the (0,2) moduli space and sometimes globally,
reducing to the (a,c) and (c,c) rings at (2,2) loci and to classical sheaf
cohomology rings at large radius. They thus define a quantum
deformation of sheaf cohomology.

Our argument begins by defining the set of (a,c) and (c,c) operators away from (2,2) loci.
The familiar definition of the (a,c) and (c,c) operators as the cohomology of left and right
moving supercharges clearly does not generalize.  Happily, a familiar stratagem from
Hodge theory suggests a natural definition which {\it does} generalize: since, at the (2,2)
point, $\{\QLB,\QL\}=\LL_0-\half \JL_0$ and $\QL^2$=0, where $\QL$ is the left-moving
supercharge of the (2,2) point,  the kernel of $\LL_0-\half \JL_0$ is in one-to-one
correspondence with cohomology classes of $\QL$.  We thus focus attention on the set
of states satisfying $\D=\pm\half q$ within right-moving $\QR$-cohomology, to which we
refer as the A and B operators.

We then prove that the OPE of these operators is non-singular on
open patches of the bundle moduli space, and globally under
certain conditions.  The argument is quite simple. First,
quantization of worldsheet spin ensures that left- and
right-moving conformal dimensions vary in lock step as we vary
bundle moduli\foot{We will assume throughout that the spectrum
varies smoothly as we vary bundle moduli - importantly, this is
the case at generic points in the moduli space of good string
compactifications. At branch points of the moduli space, the story
is modified, as we shall discuss below.}, ie $\D-\DB=n\in\IZ$.  As
a result, the holomorphic (\ie left-moving) dimension of a
right-chiral operator in a completely generic (0,2) model is
bounded from below despite the absence of a left-moving BPS bound.
By working around special points in moduli space where quantum
corrections may be controlled, \eg (2,2) loci or large radius, we
will be able to forbid singular terms in the OPEs of A or B ring
operators for finite motions in moduli space, ensuring that their
correlators are independent of insertion points and that the A and
B rings close under OPE.  Local results in hand, the left-moving
U(1) current-algebra provides global statements: in the case of
(0,2) CFTs with bundles of rank less than 8, unitarity will
actually forbid the appearance of singular terms globally on the
moduli space; when the rank is 8 or greater, it puts powerful
constraints on the form such operators must take.

It is important to emphasize that these arguments rely on (0,2)
supersymmetry and the existence of a (possibly anomalous) left-moving U(1)
current algebra, but not on sigma model perturbation theory - they
are exact to all orders and non-perturbatively in the sigma model
coupling.  In particular, these rings are not violated by worldsheet
instantons.  This fits nicely with conjectures for the
quantum cohomology of (0,2) models derived via mirror symmetry\AdamsZY, in
which both perturbative and worldsheet instanton contributions
respected the ring structure of the A and B models.

We begin in Section 2 with a review of the salient
features of $\N$=2 chiral rings in A and B twisted (2,2) models.
In Section 3 we define the A and B twists of (0,2) theories and use the results of Section 2
to identify two subsets of the right-chiral ring which are natural generalizations of the (a,c)
and (c,c) operators. In Section 4 we identify necessary and sufficient conditions for these
operators to form sub-rings of the (0,2) right-chiral ring, both in the neighborhood
of large-radius or (2,2) loci and globally on the bundle moduli space,
for both conformal and massive models.
Section 5 discusses some examples of these (0,2) rings in models of philosophical and perhaps phenomenological interest.
We close in Section 6 with
an interpretive dance.

\newsec{Chiral Rings in (2,2) Models}

We begin our study of chiral rings in (0,2) models by reviewing the special case
of (2,2) models in which, after twisting, the A and B model rings may be identified
with the cohomology of right and left moving scalar supercharges.  This is not
meant to be a self-contained introduction to $\N$=2 rings in (2,2) theories, for
which we direct the interested reader to the very beautiful original work of Lerche,
Vafa and Warner \LercheUY\ or the discussion in \WittenZZ, but simply a review
of the salient features emphasizing points that will be useful in what follows.

\subsec{(2,2) Supersymmetry}

The generators of the (2,2) superconformal algebra include left- and right-moving
supercharges $\QL$ and $\QR$ charged under left- and right-moving U(1)
R-currents $\JL$ and $\JR$, as well as holomorphic and antiholomorphic stress
tensors $\TL(z)$ and $\TR(\zb)$.  For simplicity, we focus on the right-moving
symmetry algebra; the left-moving sector is identical.  The commutators of the
right-moving algebra include
$$
\QR^2=\QR^{\dagger 2}=0, \quad \{\QR,\QRB\}=2i\pzb, \quad [\JR,\QR] = \QR, \quad [\JR,\QRB] = -\QRB .
$$
Expanding in modes, the right-moving supercommutator may be written,
$$
\{\GR_{(r)},\GRB_{(s)}\}=2\LR_{(r+s)} - (r-s)\JR_{(r+s)}  +{\overline{c}\over 3}(r^2-{1\over 4})\delta_{r+s,0},
$$
where $\overline{c}$ is the right-moving central charge, leading to a BPS bound,
$$
\DB \geq \half \qb.
$$
Operators saturate this right-moving BPS bound iff they are chiral
primary,\ie iff they satisfy $\QRB\O=0$; as a result, any product
of chiral operators is again chiral.  Similarly, operators
satisfying $\QR\O=0$ saturate the BPS bound with opposite sign,\ie
$\DB\geq\half|\qb|$.  Identical considerations apply to the
left-moving sector.  Operators which are both left- and right-BPS
form a particularly interesting set of operators; up to complex
conjugation, there are two distinct sets, the
left-chiral-right-chiral (c,c) operators and the
left-anti-chiral-right-chiral (a,c) operators.  Since the product
of two chiral operators is again chiral, the (c,c) and (a,c)
operators form rings, which have many beautiful properties which
are lovingly explored elsewhere (see \eg \MorrisonFR).

\subsec{The Twisted Models}

Twisting allows us to identify a consistent truncation of an $\N$=2 theory to its chiral
ring.  Consider again the right-moving $\N$=2. Instead of organizing operators into
representations of the Euclidean Lorentz symmetry, $U(1)_E$, we work in
representations of the diagonal subgroup of Lorentz and R symmetries,
$U(1)_E\times U(1)_R$.  This has the effect of labeling operators not by their
dimension, $\DB$, but by their dimension minus half their R-charge,
$\hb=\DB-\half \qb$; this is the dimension you would measure with a "twisted"
stress tensor, $\TR^t=\TR-\half\p_{\zb} \JR$.  Since chiral operators have twisted
dimension zero, the chiral ring of the untwisted theory becomes the ground ring
of the twisted theory, a useful simplification.

By the same token, the chiral supercharge $\QRB$ has twisted
dimension zero and transforms under $\TR^t$ as a scalar; we may
thus consistently truncate the theory by restricting to its
cohomology.  Working in $\QRB$-cohomology restricts us to the the
ground ring of the twisted model,\ie to the chiral ring of the
untwisted theory.  We could, of course, have chosen to twist by
$+\half\p_{\zb}\JR$, isolating the anti-chiral ring - however,
these twists are related by complex conjugation, and thus give
isomorphic rings.

So far we have been discussing only the right-moving symmetry algebra; in (2,2)
models, identical considerations obtain for the left-movers.  We now have two
inequivalent choices for the currents by which to twist, $J_R\mp J_L$.  In general
these give non-isomorphic models: the ground ring of the (-) twisted "A" model is
the the (a,c) ring of the untwisted (2,2) theory, while the ground ring of the (+)
twisted "B" model is the (c,c) ring.

In the special case of NLSMs, twisting has a nice geometric
meaning.  In addition to scalars $\phi^i$ coordinatizing the
target space, $X$, (2,2) NLSMs contain right and left moving
fermions\foot{This churlish notation differs from standard (2,2)
notation in treating left and right moving fermions
asymmetrically; while this obscures the full (2,2) supersymmetry,
it will prove convenient in our study of intrinsically (0,2)
models below.} coupling to the tangent bundle, $\rho^i$ and
$\l^i$, a right mover coupling to the complex conjugate of the
tangent bundle, and a left-moving fermion coupling to the
cotangent bundle, $\l_a$,\ie
$$\eqalign{
\l^i         ~\in~~ \Gamma\left(\sqrt{K_\Sigma} \otimes \phi^* T_X \right)
&\quad\quad\quad \rho^i     ~\in~~ \Gamma\left(\sqrt{\overline{K}_\Sigma}\otimes \phi^*T_X\right) \cr
\l_i        ~ \in~~ \Gamma\left(\sqrt{K_\Sigma} \otimes \phi^* T_X^* \right)
&\quad\quad\quad \rho^\ib  ~\in~~ \Gamma\left(\sqrt{\overline{K}_\Sigma}\otimes \phi^*\overline{T}_X\right)
}$$
Twisting reorganizes these fermions into representations of the twisted Lorentz algebra; for example, in the A twist,
$$\eqalign{
\l^i         ~\in~~ \Gamma\left(\phi^* T_X \right) \quad\quad\quad
\! &\quad\quad\quad \rho^i     ~\in~~
\Gamma\left(\overline{K}_\Sigma \otimes \phi^*T_X\right) \cr \l_i
~ \in~~ \Gamma\left({K}_\Sigma \otimes \phi^* T_X^* \right)
&\quad\quad\quad \rho^\ib  ~\in~~
\Gamma\left(\phi^*\overline{T}_X\right) }$$ Operators in the
twisted theory are mapped to forms in the target space, with left-
and right-moving fermion number giving holomorphic and
anti-holomorphic grading,\ie
$$\O_{p,q}\leftrightarrow\Omega^{p,q}(X).$$
Moreover, since the supercharges act as \eg $\QRB \sim
\rho^\ib{\delta\over \delta \phi^\ib}$, they push forward to the
Dolbeault operators of the target space,
$$\QRB \leftrightarrow\overline{\p}$$
The cohomology of the supercharges in the A twisted theory thus maps to the Dolbeault
cohomology of the target,
$$[\O_{p,q}] \leftrightarrow H^{p,q}(X).$$
The (a,c) ring is thus the Dolbeault cohomology ring of the target.

Of course, everything we've just said was classical - what of quantum corrections in
the NLSM?  $\N$=2 forbids any perturbative corrections; however, instanton effects
may well modify the classical correlation functions. The instanton-corrected correlators
of the twisted model thus define a quantum version of the Dolbeault cohomology,
reducing to the classical cohomology theory in the large-radius limit, where worldsheet
instanton contributions vanish.

\newsec{The A and B Operators in (0,2) Theories}

Our aim is to identify topological rings in (0,2) models which deform to the (a,c) and (c,c) rings at (2,2) loci.  As such, it is useful to observe that, since the definition of the A and B twists makes no reference to supersymmetry but only to non-anomalous left- and right-moving U(1) currents by which to deform the Lorentz generators, any (0,2) model with a good left-moving U(1) current admits the A (and, if $c_1(X)=c_1(\V)=0$, B) twist as defined above.

Of course, the (a,c) and (c,c) operators of (2,2) theories are conventionally defined as those annihilated by the appropriate right {\it and} left moving supercharges, as reviewed above; in the twisted models, they correspond to the cohomology of the left and right moving scalar supercharges.  This definition clearly does {\it not} generalize to (0,2) models, in which the left-moving supercharges are entirely absent, leaving us with a single right-moving scalar supercharge whose cohomology is the infinite dimensional space of right-chiral operators - while a beautiful mathematical object in its own right, related to the elliptic cohomology of $X$,
this is not the ring we're looking for\foot{In fact, while this work was being completed, two
preprints addressing precisely this topic appeared on the arXiv
\KapustinPT\WittenPX; in particular, both texts seek to provide a physical
interpretation of the relatively well-developed mathematical theory of chiral de Rham
operators in terms of the full $\QRB$-cohomology in the A-twist of (2,2) model or half-twist of a (0,1) or (0,2) theory, respectively.  Our interest differs from these extremely interesting papers both in studying finite-dimensional sub-rings sharing many of the properties of the familiar (a,c) or (c,c) ring, and in studying (0,2) models with left-moving fermions coupling to interesting vector bundles.}.

There is, however, an alternate definition of the (a,c) and (c,c) operators which {\it does} generalize, as mentioned above.  The basic strategy is to pullback to the worldsheet the usual Hodge Theory relation between the Dolbeault (or sheaf) cohomology, $H^{p}(X,\wedge^{q}T_X^*)$, and the zero eigenspace of the elliptic operator $\overline{\p}^{\dagger}\overline{\p}+\overline{\p}\overline{\p}^{\dagger}$, \ie the set of harmonic (p,q)-forms: since (after twisting) ${\QLB}^2=0$ and $\{\QLB,\QL\}=\LL^t_0$, states of $\LL^t_0=0$ are in one to one correspondence with $\QLB$-cohomology classes.
We may thus define the set of (a,c) and (c,c) operators in twisted (2,2) theories as the sub-set of $\QRB$-cohomology satisfying $h=\D-\half |q|=0$. While equivalent to the conventional definition of (a,c) and (c,c) operators at (2,2) points, this definition generalizes naturally to any (0,2) model with a conserved left-moving U(1) by which we can twist the
left-movers\foot{In what we hope will be a forgivable abuse of terminology, we will refer to these as the A and B operators, and the rings which they may (or may not) form as the A and B rings.
}.Quantum mechanically, while these operators clearly form a sub{\it space}, it's not entirely obvious that they form a sub{\it ring}. Proving that will be the task of the next section; the task of the remainder of this section will be to make our definition precise.  We begin by reviewing some details of (0,2) non-linear sigma-models, which will be our main examples throughout.

\subsec{The $\sigma$-model}

While our basic results will obtain for all suitably well-behaved (0,2) CFTs, we will use (0,2) NLSMs as the basic example throughout; it will thus be helpful in what follows to describe these models in some detail, making explicit use of the supersymmetry generator which will become the scalar supercharge after twisting.

Our generic example will be the non-linear $\s$-model on a rank-$r$ holomorphic bundle over a K\"ahler manifold, $\V\to X$.  Anomaly cancellation requires $c_{2}(\V)=c_{2}(T_X)$ and $c_{1}(\V)=c_{1}(T_X)$; as mentioned above and discussed in some detail below, we will actually require the slightly stronger constraint,
$$\wedge^{r}\V = K_{X}^{*}.$$
This ensures the existence of a natural inner product on $H^*(X,\wedge^*\V)$.

The fields of the NLSM include coordinates $\phi^i$ on $X$, their right-moving fermionic superpartners $\rho^i$ which couple to the tangent bundle $T_X$, and left-moving fermions
$\l^a$ (plus their auxiliary superpartners, $l^a$) which couple to the holomorphic vector bundle $\V$,
$$\eqalign{
\l^a         ~\in~~ \Gamma\left(\sqrt{{K}_\Sigma}\otimes \phi^* \V \right) ~~\!
&\quad\quad\quad \rho^i     ~\in~~ \Gamma\left(\sqrt{\overline{K}_\Sigma}\otimes \phi^*T_X\right) \cr
\l_a        ~ \in~~ \Gamma\left(\sqrt{{K}_\Sigma}\otimes \phi^* \V^* \right)
&\quad\quad\quad \rho^\ib  ~\in~~ \Gamma\left(\sqrt{\overline{K}_\Sigma}\otimes \phi^*\overline{T}_X \right)
}$$
and mix under the right-moving supersymmetry, $\QRB$, as\foot{The second supersymmetry, $\QR$, is
\eqn\secondsusy{\eqalign{
\tilde\delta\phi^{i}=\rho^{i},\quad&\qquad \tilde\delta\lambda^{a}=l^{a}+A^{a}{}_{bi}\lambda^{b}\rho^{i}\cr
\tilde\delta\rho^{i}=\ 0,\quad&\qquad \tilde\delta l^{a}=-A^{a}{}_{bi}l^{b}\rho^{i}\cr
\tilde\delta\phi^{\overline{\imath}}=\ 0,\quad&\qquad \tilde\delta\lambda_{a}=0\cr
\tilde\delta\rho^{\overline{\imath}}=\partial_{\overline{z}}\phi^{\overline{\imath}},&\qquad \tilde\delta l_{a}=\partial_{\overline{z}}\lambda_{a}
}}
The supersymmetry algebra is satisfied, provided the (2,0) part of the curvature vanishes:
$$
A^{a}{}_{b[i,j]}-A^{a}{}_{c[i}A^{c}{}_{bj]}=0
$$
Note that we have introduced a shift in the definition of the $l^{a}$, so as to make all of the $\QRB$ supersymmetry variations gauge-covariant. This greatly simplifies many formul\ae\ in the twisted model. In the untwisted model, one might prefer a more symmetrical choice. }
\eqn\Qnilp{\eqalign{
\delta\phi^{\overline{\imath}}=\rho^{\overline{\imath}},&\qquad \delta\phi^{i}=\ 0\cr
\delta\rho^{\overline{\imath}}=\ 0,&\qquad \delta\rho^{i}=\partial_{\overline{z}}\phi^{i}\cr
\delta\lambda_{a}=l_{a},&\qquad \delta\lambda^{a}=\ 0\cr
\delta l_{a}=\ 0,&\qquad \delta l^{a}=D_{\overline{z}}\lambda^{a}+F^{a}{}_{bi\overline{j}}(\phi)\lambda^{b}\rho^{i}\rho^{\overline{\jmath}},
}}
where $F^{a}{}_{bi\overline{\jmath}}=-A^{a}{}_{bi,{\overline{\jmath}}}$ is the curvature form of $\V$ and
$$
 D_{\overline{z}}\lambda^{a}= \partial_{\overline{z}}\lambda^{a}+A^{a}{}_{bi}(\phi)\partial_{\overline{z}}\phi^{i}\lambda^{b}.
$$
The action can then be written as
\eqn\treeAction{
  S = \int d^{2}z\ \{ \QRB, \chi\} + S_{\omega}
}
where
\eqn\treeB{
  \chi = g_{\overline{\imath}j}(\phi)\p_z\phi^{\overline{\imath}}\rho^{j} +\lambda_{a}l^{a}
}
and
\eqn\somega{
 S_{\omega}= {1\over2}\int d^{2}z\ g_{\overline{\imath}j}(\phi)(\partial_{\overline{z}}\phi^{\overline{\imath}}\partial_{z}\phi^{j}-\partial_{z}\phi^{\overline{\imath}}\partial_{\overline{z}}\phi^{j}) + i \int \phi^{*}B
}
is ($i$ times) the pullback of the complexified K\"ahler form, $B+i\omega$. (To avoid the concomitant complexities, we  take the 2-form, $B$, to be closed). In its full component glory, then,
\eqn\fullAction{\eqalign{
 S=& \int d^{2}z\  {1\over2}g_{\overline{\imath}j}(\phi)
 (\partial_{z}\phi^{\overline{\imath}}\partial_{\overline{z}}\phi^{j}
 +\partial_{z}\phi^{j}\partial_{\overline{z}}\phi^{\overline{\imath}})
 - g_{\overline{\imath}j}(\phi)
 \rho^{j}(\delta^{\overline{\imath}}_{\overline{k}}\partial_{z}
 +\Gamma^{\overline{\imath}}_{\overline{k}\overline{l}}(\phi)\partial_{z}\phi^{\overline{l}})\rho^{\overline{k}}\cr
 & + \lambda_{a}D_{\overline{z}}\lambda^{a} +F^{a}{}_{bi\overline{\jmath}}(\phi)\lambda_{a}\lambda^{b}\rho^{i}\rho^{\overline{\jmath}}+l_{a}l^{a} + i\int \phi^{*}B
}}

Classically, this model possesses a right-moving R-symmetry and a left-moving flavour symmetry forming a $U(1)_R\times U(1)_L$ global symmetry group under which $\rho^i$ and $\rho^\ib$ have charges $(\pm 1,0)$,  $\lambda^{a}$ and $\lambda_{a}$ have charges $(0,\pm 1)$, and $l^a$ and $l_a$ have charges $(\pm1,\pm1)$.  Classically, then, we may shift the spins of all fields by a linear combination of their charges; in the model twisted by $J=(1-2s)J_L+(2\sb-1)J_R$, the ``fermions" transform as
\eqn\fermTransf{\eqalign{
\l^{a}   ~\in~~ \Gamma\left(K_\Sigma^{(1-s)}\otimes \phi^*{\V}\right) \qquad ~~ &\qquad
\r^{i}    ~\in~~ \Gamma\left(\overline{K}_\Sigma^{\overline{s}}\otimes \phi^*T_X\right) \cr
\l_{a}   ~ \in~~ \Gamma\left({K}_\Sigma^{s}\otimes \phi^*\V^{*}\right) \qquad\qquad\, &\qquad
\r^{\ib} ~\in~~ \Gamma\left(\overline{K}_\Sigma^{(1-\overline{s})}\otimes \phi^*\overline{T}_X\right).
}}
Here, $s$ and $\sb$ label the spin of the left- and right-moving fermions; the untwisted theory has $s=\sb=\half$. The auxiliary bosonic fields transform as
\eqn\auxTransf{
~~ \qquad l^{a}    ~\in~~ \Gamma\left(K_\Sigma^{(1-s)}\otimes \overline{K}_{\Sigma}^{\overline{s}}\otimes \phi^*{\V}\right)  \quad \qquad
l_{a}    ~\in~~ \Gamma\left({K}_\Sigma^{s}\otimes \overline{K}_{\Sigma}^{(1-\sb)}\otimes \phi^*\V^{*}\right).
}

Quantum-mechanically, these $U(1)$ symmetries are anomalous, with the charge-violation on a genus-$g$ surface given by
$$\eqalign{
 \delta q_{L}&=(1-g)(1-2s)r + \phi^{*}c_{1}(\V)\cr
 \delta q_{R}&=(1-g)(2\overline{s}-1)d + \phi^{*}c_{1}(T_X).
}$$ In the untwisted ($s=\sb=\half$) model, the first terms
vanish. To ensure that the twisted model has a non-anomalous
Lorentz symmetry, we should twist only by a non-anomalous
combination of global currents\foot{While we require the twisted
Lorentz symmetry to be non-anomalous, the global symmetry may, and
generally will, pick up an anomaly after twisting.}.  Since, in
the models we consider, $c_{1}(T_X)$=$c_{1}(\V)$, the current
$J_R-J_L$ is always nonanomalous, allowing us to twist by this
$U(1)$ to obtain the $\sb=s=1$ $A$-model, in which the fermions
transform as \eqn\fermTransfA{\eqalign{ \l^{a}      ~\in~~
\Gamma\left(\phi^*{\V}\right) \quad\quad\quad ~\! &\quad\quad\quad
\rho^{i} ~\in~~ \Gamma\left(\overline{K}_\Sigma \otimes
\phi^*T_X\right) \cr \l_{a} ~ \in~~ \Gamma\left({K}_\Sigma \otimes
\phi^*\V^{*}\right) &\quad\quad\quad \rho^{\overline{\imath}}
~\in~~ \Gamma\left(\phi^*T_X^*\right) }} If
$c_{1}(T_X)$=$c_{1}(\V)=0$,\ie $X$ is Calabi-Yau, $J_R$ and $J_L$
are separately nonanomalous and other twists are possible.  For
instance, the $s=0$, $\sb=1$ $B$-model involves twisting by
$J_L+J_R$, while the $s=\half$, $\sb=1$ half-twisted model
involves twisting by $J_R$ alone\foot{Note that the difference
between the various twisted models is less dramatic in (0,2) than
in (2,2) theories - in particular, {\it all} models are subject to
worldsheet instanton corrections. That said, exchanging the roles
of $\l_{a}$ and $\l^{a}$ while reversing the sign of $J_{L}$ (and
changing the $l^a$-dependence of \secondsusy\Qnilp, which is
trivial on-shell) maps $A(X,\V)$ into $B(X,\V^*)$, imposing
interesting constraints on the form of instanton corrections.}.

Even in the non-Calabi-Yau case, we might be tempted to consider these other twists, or relax the condition $\wedge^{r}\V= K_{X}^{*}$. In doing so, however, we pay a price (in addition to giving up the existence of a trace on the algebra, as discussed below): while the local physics of these more general theories looks fairly familiar in $\sigma$-model perturbation theory, the $U(1)$ by which we twist is almost invariably violated by worldsheet instantons, changing the physics radically\foot{On a flat worldsheet, the theory, \fullAction, suffers from a $\sigma$-model anomaly, unless
$$ch_{2}(\V)-ch_{2}(T_X)=0\ .$$
On a curved worldsheet, there is an additional contribution the anomaly 4-form,
$$
 {1\over 2} c_{1}(\Sigma) \bigl((2s-1)c_{1}(\V)-(2\overline{s}-1)c_{1}(T_X)\bigr)
$$
The untwisted theory ($s=\overline{s}=1/2$) is non-anomalous, but twisted theories are only sensible when this quantity vanishes.}.  Consider, for instance, the $\IP^{1}$ model, with $\V=0$, discussed in \WittenPX. In $\sigma$-model perturbation theory, there is a rich spectrum of operators in the $\QRB$-cohomology (provided one doesn't restrict oneself to scaling dimension zero). However, worldsheet instantons correct the $\QRB$-action in such a way that all the operators pair up, and the $\QRB$-cohomology of the exact theory is empty.

Aside from the existence of other twists, there's another distinction between the Calabi-Yau case and more general ``massive'' $\sigma$-models with $c_1\neq0$ which will be very important for us below. At the classical level, both are conformally-invariant: $T_{z\overline{z}}= 0$ and the other components of the stress tensor (for the $A$-model),
\eqn\treeStress{\eqalign{
  T_{zz}&= -g_{i\overline{\jmath}}(\phi) \partial_{z}\phi^{i}\partial_{z}\phi^{\overline{\jmath}}- \lambda_{a}D_{z}\lambda^{a}\cr
  T_{\overline{z}\overline{z}}&= -g_{i\overline{\jmath}}(\phi) \partial_{\overline{z}}\phi^{i}\partial_{\overline{z}}\phi^{\overline{\jmath}}+g_{i\overline{\jmath}}(\phi) \rho^{i}(\partial_{\overline{z}
}\rho^{\overline{\jmath}}+\Gamma^{\overline{\jmath}}_{\overline{k}\overline{l}}\partial_{\overline{z}}\phi^{\overline{k}}\rho^{\overline{l}})
}}
satisfy
\eqn\treeSatisfy{\eqalign{
 [\QRB,T_{zz}]&= -l_{a}D_{z}\lambda^{a} + \Bigl(-g_{i\overline{\jmath}}(\partial_{z}\rho^{\overline{\jmath}}+\Gamma^{\overline{\jmath}}_{\overline{k}\overline{l}}\partial_{z}\phi^{\overline{k}}\rho^{\overline{l}})+F^{a}{}_{bi\overline{\jmath}}\lambda_{a}\lambda^{b}\rho^{\overline{\jmath}}\Bigr)\partial_{z}\phi^{i}\cr
 &= 0\ \hbox{\rm on-shell}\cr
 T_{\overline{z}\overline{z}}&= \{\QRB, -g_{i\overline{\jmath}}\rho^{i}\partial_{\overline{z}}\phi^{\overline{\jmath}}\}.
}}
Thus $T_{\overline{z}\overline{z}}=0$ in $\QRB$-cohomology, while $T_{zz}$ descends to an operator on the $\QRB$-cohomology.

The fact that $T_{zz}$ is not $\QRB$-exact, even classically, means that the (0,2) $A$-model is a 2D {\it conformal} field theory, rather than a 2D {\it topological} field theory. Our interest in the ground ring is that it forms a ``topological subsector'' of this conformal field theory.

Quantum mechanically, the conformal structure is violated by the one-loop $\beta$-function. Renormalization adds to the action a term of the form,
$$
 \Delta \chi_{\rm 1-loop}= \kappa_{1}\ R_{\overline{\imath}j} \partial_{z}\phi^{\overline{\imath}}\rho^{j}+\kappa_{2}\ g^{\ib j}F^{a}{}_{b \ib j}\lambda_{a}l^{b}
$$
for some divergent constants $\kappa_{{1,2}}$ (with a concomitant shift of $S_{\omega}$ by the pullback of the Ricci form). In the Calabi-Yau case, we can make the conformal anomaly {\it vanish} by choosing the Ricci-flat metric and a solution to the Uhlenbeck-Yau equation, $g^{\ib j}F^{a}{}_{b \ib j}=0$. In the ``massive models'', however, conformal invariance is necessarily lost, and there {\it is} nontrivial RG running. It is, however, $\QRB$-trivial, and so does not affect the correlation functions of operators in the $\QRB$-cohomology\foot{In perturbation theory, that statement was precisely correct. At the level of worldsheet instantons, the result is to trade the exponential of the pullback of the K\"ahler form for the dimensionful scale, via dimensional transmutation.}.  More precisely, $T_{z\zb}$, while no-longer vanishing, remains $\QRB$-exact,
$$
  T_{z\overline{z}} \propto \{ \QRB, \Delta G_{z\zb}\}
$$
and $T_{\overline{z}\overline{z}}$ remains $\QRB$-exact, so, on the level of the $\QRB$-cohomology, we are in almost as good shape as before.
However, there is a fly in the ointment. Back in \treeSatisfy, we noted that, classically, $[\QRB, T_{zz}]$ closed onto the equations of motion. In the massive $A$-model, this fails quantum mechanically:
\eqn\QClosed{
[\QRB, T_{zz}]=\p_z V\neq 0
}
where $V=R_{\ib j}\p_z\phi^j\r^\ib+...$ and its derivative does {\it not} vanish by the equations of motion.  As a result, general changes of holomorphic coordinate do {\it not} preserve $\QRB$-cohomology classes, so the A-model is {\it not} conformal, though conservation of the stress tensor, $\p_\zb T_{zz} = - \p_z T_{z\zb}$, does ensure that the left-moving stress tensor is holomorphic up to $\QRB$-trivial terms, $\p_\zb T_{zz} \sim 0$.

Fortunately, our arguments do not depend on full conformal invariance; as we shall see in the next section, all we really need is that holomorphic scaling dimension and momentum remain good quantum numbers in $\QRB$-cohomology, \ie\ that $\LL_0$ and $\LL_{-1}$ commute\foot{Since $\p_z T_{z\zb}$ is nonzero, $T_{zz}$ is not holomorphic so it does not make sense to speak of its Laurent coefficients, $L_n$. However, since $T_{z\zb}$ is
$\QRB$-exact, so is the non-holomorphic dependence of $T_{zz}$. Thus, it does make sense to talk about the Laurent coefficients, $L_n$, when working modulo $\QRB$-exact operators.} with $\QRB$. This is easily verified. For example, \QClosed\ and the holomorphy of $T_{zz}$ imply $[\QRB,\LL_{-1}]=0$. Similarly, since $\QRB$ is by construction spinless after twisting,
$$
0=[S,\QRB]=i[\LL_0,\QRB]-i[\LR_0,\QRB] = i[\LL_0,\QRB],
$$
as $\LR_0$ is a $\QRB$-commutator (since $T_{\zb\zb}$ is).
$\LL_0$ thus preserves cohomology class.
This is enough for our purposes.

Summing up the last two paragraphs, the massive A-model is a holomorphic field theory invariant under global dilatations and translations, though not general holomorphic coordinate transformations. As we shall see, this provides just enough control
to ensure the existence of the A and B rings in massive (0,2) models, at least in open balls around special points in the moduli space.  We will return to this subtlety in our discussion of massive (0,2) models in the next section; for now we will restrict attention to models which were already conformal {\it before} twisting.

\subsec{Bundle Moduli}
It is instructive to work out the ``integrated vertex operators'' which represent infinitesimal deformations of the moduli of the bundle of our (0,2) $\sigma$-model. Infinitesimal deformations of the holomorphic structure of the vector bundle, $\V$, correspond to elements $h\in H^{1}(X, {\rm End}\V)$. Explicitly, these are $h^{a}{}_{b\overline{\imath}}(\phi)$, which are $\overline{\partial}$-closed and traceless,
$$
  h^{a}{}_{b[\overline{\imath},\overline{\jmath}]}
  = \delta^{b}{}_{a}h^{a}{}_{b\overline{\imath}}=0
$$
modulo those which are $\overline{\partial}$-exact. We can write a vertex operator,
\eqn\holostrV{
V = [h^{a}{}_{b\overline{\imath}}\partial_{\overline{z}}\phi^{\overline{\imath}} + h^{a}{}_{b\overline{\imath};j}\rho^{j}\rho^{\overline{\imath}}]\lambda_{a}\lambda^{b} + h^{a}{}_{b\overline{\imath}}\rho^{\overline{\imath}}\lambda_{a}l^{b}
}
where
$$
 h^{a}{}_{b\overline{\imath};j} = h^{a}{}_{b\overline{\imath},j}+ A^{a}{}_{cj}h^{c}{}_{b\overline{\imath}}- h^{a}{}_{c\overline{\imath}}A^{c}{}_{bj}
$$
This represents a deformation of holomorphic structure of $\V$, which --- when referred to the original basis of local sections, $\lambda^{a}$ --- adds a (0,1) component to the connection, while preserving the fact that the curvature is of type (1,1) and preserving its trace. The second term in \holostrV, involving the auxiliary field $l^{a}$, was added for convenience; it vanishes on-shell. If we demand that the curvature of the deformed bundle satisfy the Uhlenbeck-Yau equation, we should choose a harmonic representative, $g^{\overline{\imath}j}h^{a}{}_{b\overline{\imath};j}=0$, for this cohomology class.

A short computation shows that
$$
 [\QRB,V] = \partial_{\overline{z}}(h^{a}{}_{b\overline{\imath}}\rho^{\overline{\imath}}\lambda_{a}\lambda^{b})+\dots
$$
where $\dots$ are terms proportional to
$$\partial_{\overline{z}}\lambda_{a}-A^{c}{}_{aj}\lambda_{c}\partial_{\overline{z}}\phi^{j}-F^{c}{}_{aj\overline{k}}\lambda_{c}\rho^{j}\rho^{\overline{k}}
$$
and to $l_{a}$, both of which vanish on-shell. So $\int d^{2}z V$ is a $\QRB$-invariant deformation of the $\sigma$-model.

As with deformations of the holomorphic structure of $\V$, infinitesimal deformations of the complex structure can be viewed as deformations of the $\overline{\partial}$ operator on $X$ which preserve $\overline{\partial}^{2}=\{\overline{\partial},\partial\}=0$.
Such a deformation can be written as $\overline{\partial}\to \overline{\partial} + d\phi^{\overline{\imath}}h_{\overline{\imath}}{}^{j}\partial_{j}$, where $h_{\overline{\imath}}{}^{j}(\phi)\in H^{1}(X,T_{X})$.

The vertex operator which implements this is somewhat lengthy to write down in full. It can most succinctly be written as
\eqn\complexdefo{
 V= \tilde{\delta}U,\quad \hbox{where}\ U= h_{\overline{k}}{}^{j}(g_{\overline{\imath}j}\partial_{z}\phi^{\overline{\imath}}-F^{a}{}_{bj\overline{m}}l_{a}\lambda^{b}\rho^{\overline{m}})\rho^{\overline{k}}
}
using the second supersymmetry, \secondsusy. By construction, we have $\{\QRB, V\} =\partial_{\zb}U + ...$, where ``...'' are terms which vanish by the equations of motion.

Finally, deformations of the complexified K\"ahler structure represent another set of $\QRB$-invariant deformations of the action which are even easier to understand. As we saw, the dependence of the action \treeAction\ on the complexified K\"ahler class is given, up to $\QRB$-trivial terms, by $S_{\omega}$ \somega. Shifting $B+i\omega$ by a complex, closed (1,1)-form, $b$, shifts $S$ by $i\int\phi^{*}b$.

In the (2,2) case, A-model correlation functions are independent of the complex structure moduli, while B-model correlators, which do not receive world-sheet instanton corrections, are independent of the K\"ahler moduli.
In the (0,2) context, the story is, {\it a-priori}, more complicated: to begin, we have a third class of moduli, the deformations of the holomorphic vector bundle, $\V$, on which correlators may depend; more troublingly, both A {\it and} B twisted models now receive instanton corrections, and thus depend on both K\"ahler and complex structure moduli, as well as the bundle moduli. One powerful constraint, that we can see from the explicit construction of the deformations, is that the dependence on these moduli is {\it holomorphic}. Further restrictions arise for (0,2) models in which $\V$ is a deformation of the tangent bundle, as they must reproduce the familiar results at the (2,2) locus. However, away from such loci, or in general (0,2) theories without (2,2) loci on their moduli spaces, simplifications appear few and far between.

That said, explicit computations often reveal that the most general possible dependence on the moduli does not, in fact, arise. Rather, one finds intriguing hints of various ``non-renormalization'' theorems ensuring that the ring relations remain independent of certain moduli. The full implications of these observation are, however, beyond the scope of the present paper.

\subsec{Right-chiral ground states in (0,2) models}

Operators in the twisted theory are mapped to bundle-valued forms in the target space. We will drop the auxiliary fields to simplify the notation, as they won't contribute to correlation functions at non-coincident points, their propagators being trivial.

Since our single scalar supercharge acts as $\QRB \sim \r^\ib{\delta\over \delta \phi^\ib}$, operators in $\QRB$-cohomology take the form,
$$
  \O_{\ib_1 \dots \ib_p} \r^{\ib_1} \dots \r^{\ib_p}
$$
with
$$
\O_{\ib_1 \dots \ib_p}= \O_{\ib_1 \dots \ib_p}(\phi,\partial_{z}\phi,\partial_{z}^{2}\phi,\dots;\overline{\phi},\partial_{z}\overline{\phi},\partial_{z}^{2}\overline{\phi},\dots;\l^{a},\partial_{z}\l^{a},\dots; \l_{a},\partial_{z}\l_{a},\dots)
$$
where we've taken the liberty of using the equations of motion for $\rho^{\overline{\imath}}$ to trade $z$-derivatives of $\rho^{\overline{\imath}}$ for the other fields and their derivatives. $\QRB$-closedness implies holomorphy in (the constant mode of) $\phi$, but is otherwise not very contraining.

Restricting to operators with $L^t_{0}=0$ simplifies this structure dramatically: since $\p_{z}^{k}\phi$, $\p_{z}^{k}\overline{\phi}$ and $\l_{a}$ all contribute positively to the (twisted) dimension, operators of dimension $h=0$ take the beautifully simple form
$$
\O_{{p,q}} \sim  \O_{\ib_1\dots\ib_p;a_1\dots a_q}(\phi) \r^{\overline{\imath}_{1}}\dots\r^{\overline{\imath}_{p}} \l^{a_{1}}\dots\l^{a_{q}}
$$
Modding out by $\QRB$-trivial operators, these are in 1-to-1 correspondence with elements of the sheaf cohomology,
$$\O_{p,q}\in H^{p}(X,\wedge^q{\V}^{*}).$$
Serre duality then provides a trace on the ring iff the dualizing sheaf $\wedge^r\V\otimes K_X$ is trivial:
$$\eqalign{
H^p(X,\wedge^q{\V}^{*}) &=
H^{d-p}(X,\wedge^{r-q}\V^*\otimes \wedge^r\V\otimes K_X)^* \cr
&= H^{d-p}(X,\wedge^{r-q}\V^{*})^*.
}$$
Of course, $\wedge^r\V\equiv K^*_X$ implies $c_1(T_X)=c_1(\V)$, which was already required to have a nonanomalous left-moving $U(1)$ by which to twist; it's pleasing that this slightly stronger condition also guarantees (classically) the existence of a trace on our ring.

\subsec{Correlators of $\QRB$-Cohomology Classes in the Twisted Models}

Correlators of right-chiral operators in the twisted model satisfy several very important properties which will be crucial in what follows.
First, since the twisted vacua are annihilated by the supercharges, correlators including an insertion of a $\QRB$ commutator vanish,
$$
\langle\O_1...\{\QRB,M\}...\O_s\rangle=0.
$$
Since the right-moving stress tensor is trivial in $\QRB$-cohomology, $\TR=\{\QRB,\GR\}$, correlators of $\QRB$-chiral operator with insertions of the stress tensor automatically vanish,
$$
\langle \TR \prod_i\O(z_i,\zb_i)\rangle ~ = ~ \langle  \{\QRB,\GR\} \prod_i\O(z_i,\zb_i)\rangle ~ = ~ 0.
$$
Correlators of $\QRB$-chiral operators are thus completely independent of $\zb$, depending only holomorphically on their insertion points on the worldsheet,
$$
\langle\prod_i\O(z_i,\zb_i)\rangle~=~ \langle\prod_i\O(z_i)\rangle.
$$
Scaling invariance and conservation of the left-moving $U(1)$ thus ensure that the OPE of two $\QRB$-chiral operators takes the form
\eqn\oihfe{
\O_a(z)\O_b(0) = \sum_{q_c=q_a+q_b} {f_{abc}\over z^{h_a+h_b-h_c}} \O_c(0) ,
}
where $\O_c$ is necessarily $\QRB$-chiral. Now suppose one could show that there existed a subset of the $\QRB$-cohomology whose OPEs were completely non-singular,
$$
\O_a(z)\O_b(0) = \sum_{q_c=q_a+q_b} f_{abc} \O_c(0) + O(z).
$$
On good physical grounds, we do not expect the correlation function to diverge as $z\to\infty$, so it can be extended analytically to the Riemann sphere.
Since the only holomorphic function on a compact Riemann
surface without a pole is the constant function, correlators of
such magical operators would be completely independent of
insertion points, and thus of the worldsheet metric,
forming an extremely simple topological ring,
$$
\O_a\O_b=f_{abc}\O_c.
$$
As we shall see in the next section, under very mild conditions, the ground operators of A and B twisted (0,2) models introduced above satisfy precisely such a condition, their non-singular OPEs providing a ring structure and ensuring the topological character of their correlators.  Let's prove it.

\newsec{A and B Rings in (0,2) Models}

As we shall see, (0,2) superconformal symmetry together with a
left-moving U(1) current algebra satisfying simple conditions will
suffice to ensure the existence of rings (in fact, finite dimensional algebras) of topological operators
closing under non-singular OPE and forming the ground rings of the
A and B twisted models.  We begin by considering deformations of
(2,2) models, then generalize to intrinsically (0,2) SCFTs, and
finally address massive (0,2) models.

\subsubsec{Local Results for (0,2) Deformations of (2,2) Models}

Let's begin with a flanking maneuver. Consider the (c,c) ring of a (2,2) SCFT.
Left and right chirality ensures that these operators saturate both left
and right twisted BPS bounds, $h=0$ and $\hb=0$. Note
that worldsheet conformal invariance implies the quantization of
worldsheet spin, $h-\hb=s\in\IZ$.

Now deform this theory by a marginal operator which preserves both
(0,2) and the left-moving U(1) R-current,\ie a dimension (1,1)
operator with R-charge (0,2) (in the case of an NLSM, this would
correspond to an element of $H^1(X,End(T_X))$ of the form
discussed in the previous section). Denote by $\a$ the deformation
parameter. We make the (quite reasonable) assumption that the
spectrum varies smoothly under this marginal deformation,\ie that
we do not begin with or approach a singular CFT.

Since this deformation preserves worldsheet conformal invariance, $h(\a)-\overline{h}(\a)=s$ continues to hold in the deformed theory.
This allows us to translate the antiholomorphic BPS bound into an effective bound on the holomorphic weights.
Explicitly, since the dimensions and spin of operators remain well-defined and vary smoothly under our deformation, the left-moving conformal dimension of every operator in the theory is pegged to vary in
lock step with its right-moving conformal dimension,
$$\delta_\a h = \delta_\a \overline{h}.$$
The unbroken anitholomorphic BPS bound $\hb(\a)  \geq 0$ thus translates into a bound on the holomorphic dimension {\it away} from the (2,2) locus,
$$
h(\a) = \hb(\a) + s  \geq  s \in\IZ
$$
In other words, the holomorphic weight of every operator is bounded by the amount by which the operator failed to saturate the antiholomorphic BPS bound in the undeformed theory.  In particular, the holomorphic weight of a right-chiral operator cannot decrease as we turn on the deformation - and all this despite the absence of any left-moving supersymmetry!

This bears repeating.
The deformed theory does not have a holomorphic BPS bound, and
there will in general exist operators with $h(\a)<0$. However,
since conformal invariance implies that
$h(\a)-\overline{h}(\a)=s\in\IZ$, operators which were
right-chiral at the (2,2) point (\eg the (a,c) or (c,c) operators)
continue to satisfy $h(\a)\geq 0$ even away from the (2,2) locus.
By the same token, operators which were {\it not} right-chiral at
the (2,2) locus may flow down, keeping $h(\a)-\overline{h}(\a)=s$
fixed, until they saturate the right-BPS bound; this puts a bound
on the amount by which the holomorphic weight can flow, $h(\a)\geq
s$; we shall return to this possibility below.

First, though, let's study the consequences of this bound on OPEs of A-model groundstates.  Let $\{\O_a\}$ be the subset of $\QRB$-cohomology with holomorphic weight $h_a=0$, forming the groundstates of the A-model; by the above arguments, these operators continue to satisfy $h(\a)=0$ away from
the (2,2) locus.  Conformal invariance and left-moving $U(1)$ conservation thus imply
$$\eqalign{
\O_a(z)\O_b(0) &= \sum_{q_c=q_a+q_b} {f_{abc}\over
z^{h_a(\a)+h_b(\a)-h_c(\a)}} \M_c(0)\cr &= \sum_{q_c=q_a+q_b}
{f_{abc}\over z^{-h_c(\a)}} \M_c(0)  ,\cr }$$ Note that $\M_c$
does {\it not}, in general, obey $h_c=0$,\ie the OPE does {\it
not}, in general, close on the ground  operators {\it within}
$\QRB$-cohomology.  However, as discussed at the end of the last
section, to ensure that correlators of A and B operators remain
completely independent of insertion point and continue to define a
topological ring it is sufficient to show that their OPEs remain
non-singular away from the (2,2) loci.

When, then, can singular terms arise?  The appearance of poles in
the OPE requires the existence of a right-chiral operator with
$$
h_c(\a)<0,
$$
ie the ring relations can only become singular if there exists a
right chiral operator violating the erstwhile left-moving BPS bound. By
the above, this operator must have flowed from an operator which was
{\it not} right-chiral at the (2,2) locus -- otherwise it would
continue to respect the left-moving
BPS bound away from (2,2) locus, $h_c(\a)\geq 0$.

By quantization of worldsheet spin and continuity of the spectrum under marginal deformations,
this operator can only have entered the right-chiral ring after a
finite deformation away from the (2,2) locus. This ensures that
the OPEs of ground-ring elements remain closed and non-singular in at
least an open neighborhood of the (2,2) locus, proving that the
twisted ground-ring exists, as a topological ring, even away from (2,2) loci.

In fact, this argument gives us much more. In an {\it arbitrary} (0,2) $\sigma$-model, the engineering dimensions of all the operators in the twisted theory are non-negative. To find a ``dangerous'' operator, with $(h(\a),\hb(\a))=(-|s|,0)$, we need to start with an operator with non-negative engineering dimension, $(h(\a),\hb(\a))=(h,h+|s|)$, which picks up a large negative anomalous dimension under RG flow. That clearly can't happen while the $\sigma$-model remains weakly coupled. Remarkably, even far from weak coupling, there are constraints, to which we now turn.

\subsec{Global Results  for (0,2) SCFTs}

If the (0,2) model was superconformal {\it before} twisting, then unitarity of the untwisted model provides further powerful constraints. The stress tensor of a
unitary (0,2) SCFT with a $r$ left-moving fermions counted by a
left-moving U(1) current algebra can be put in Sugawara form,
$$
T = T' + {1\over 2r}:J^2(q): .
$$
Since $T'$ is the stress tensor of the (unitary!) coset conformal field theory, its spectrum of conformal weights is non-negative. Thus we have a bound relating the U(1) charge and the (untwisted) dimension,
\eqn\qunitarity{
\D \geq {q^2\over 2r} .
}

Now twist. In the A-model (B-model), the conformal weights are
$$\eqalign{
h &= \D \mp \half q \cr
\hb &= \DB + \half\qb
}$$

We are interested in operators with $\hb=0$, \ie $\DB=\half\qb$.
At the same time, we wish to drive $h=\D\mp \half q$ sufficiently
negative. But \qunitarity\ implies
$$
  h \geq {q^2\over 2r} \mp q/2 = {q(q\mp r)\over 2 r}
$$
The RHS is minimized for $q=\mp r/2$, so we have the bound
$$
  h \geq -r/8
$$
To get a pole, we need $h=s\in\IZ<0$; this requires $r\geq 8$. For $r< 8$, {\it unitarity of the untwisted theory forbids a large negative anomalous dimension}, even if the $\sigma$-model is strongly coupled.  Thus, in the absence of any candidate ``dangerous'' operators, the ground ring must persist even deep into the (0,2) moduli space.

\subsubsec{Massive (0,2) Theories}

As we saw at the end of Section $3.1$,  in A-models for ``massive'' (0,2) theories, $T_{z\zb}$ and $T_{\zb\zb}$ are $\QRB$-exact, while $\LL_0$ and $\LL_{-1}$ are $\QRB$-closed, ensuring that, for fixed but arbitrary metric, these A-models are holomorphic field theories invariant under global dilitations and translations.  Operators in $\QRB$-cohomology thus carry well-defined holomorphic scaling dimensions and spin, and are invariant under translations of their insertion points.

This is just enough to apply the arguments of the previous subsections to A-twisted massive models.  It is simple to verify that OPEs of spinless operators are again non-singular in the neighborhood or large-radius or (2,2) points, and that their correlators are regular at worldsheet infinity.  Again, operators can flow down and enter the ring in pairs, but this can only happen after some finite bundle deformation.

More explicitly, to define
$$
 \langle \CO_{1}(z_{1})\dots\CO_{n}(z_{n})\rangle
$$
we need to specify a metric; for simplicity, we will use the round metric on the sphere,
\eqn\round{
  ds^{2} = { 4|dz|^{2}\over (1+|z|^{2}/R^{2})^{2}}.
}
As we have argued, correlation functions of ground ring operators are both holomorphic\foot{Since we are using the round metric, they are also invariant under $SO(3)$ rotations of the sphere -- recall that the ground ring operators are spinless -- so they inherit an {\it accidental} $PSL(2,\IC)$ symmetry.} and invariant under dilitations, and thus independent of the parameter $R$ in the round metric \round. As before, we need to investigate the possibility of poles as $z_{i}\to z_{j}$.

The massive model has a dimensionful scale, $\Lambda$, and we have a dimensionless parameter, $\epsilon = R \Lambda$ at our disposal. At finite $\epsilon$, we need to use the curved-space Green's functions associated to the metric \round, rather than the flat space ones.
But all we are interested in is whether there is a pole as $z_{i}\to z_{j}$. This is entirely governed by the operator product expansion of $\O_i$ and $\O_j$. Again, the danger is that there exists a $\QRB$-invariant operator of spin $N$, with large negative anomalous dimension, which might contribute a singular term to the OPE. But, for small enough $\epsilon$, the theory is weakly coupled, and the anomalous dimensions are all small. Hence, there can be no such singular contribution to the correlation function. Since it is a globally holomorphic function on the sphere, it must be a constant.

\newsec{Examples}

While (0,2) ground rings have previously not been systematically investigated,
special cases have cropped up in the literature.
Here we review two examples where the rings have been explicitly
constructed, providing concrete examples of the abstract
structures discussed above.

\subsec{Explicitly Solvable Conformal Models}

Much work has gone into the exact solution of various (0,2) models,
focusing largely on Landau-Ginsburg orbifolds and their
deformations \KachruPG\DistlerMK\KachruWM\BlumenhagenEW. In these cases,
the ring structure may be extracted by inspection. In a
remarkable paper, Blumenhagen, Schimmrigk and Wi\ss kirchen (BSW)
did precisely this, identifying chiral sub-rings in a
series of $(0,2)$ deformation of Gepner models models
\BlumenhagenEW. Notably, their construction relies crucially on
the existence of a left-moving U(1) current; in retrospect, we may
twist by this U(1) and check that the resulting ground ring is
precisely the ring they identify.  The interpretation as the
cohomology ring $H^p(X,\wedge^q V)$ was also pointed out.

Consider, for example, their ``(80,0) model'', a deformed $3^{\otimes 5}$ Gepner
model with a spacetime SO(10) gauge group, whose geometric phase corresponds to a rank four
bundle over a complete intersection in the weighted projective
space $W\IP_{(1,1,1,1,2,2)}$ (the familiar two-parameter Calabi-Yau ``Example 2''
from \MorrisonFR, also studied for example in \DistlerMK). BSW explicitly
checked for the existence of the ring by calculating the massless
Yukawa couplings with the SO(10) representation structure
$\langle\bf{10\cdot 16\cdot 16}\rangle$ that define the B-ring.
The corresponding vertex operators saturate the chiral bound
$\D\geq {1 \over 2}|q|$.  The complete ring structure is given in Table 1.

BSW also checked that this ring agreed with the coordinate ring derived
from the superpotential of the associated gauged linear sigma model, \ie
$$
{\cal R}={\IC[\Phi_i]\over \{J_a(\Phi_i)=0\}}
$$
where the $W_{(0,2)}=\int\!d\t \Gamma^a J_a(\Phi_i)$ is the (0,2)
superpotential.  Since $J_a=\partial_a\! W$ at (2,2) loci,
this matches the usual B-ring at (2,2)-points; at generic points,
this is precisely the B ring of the (80,0) (0,2) theory.

\thicksize=1pt
\vskip12pt
\begintable
$K_1$ | $\langle K_1\xi_1\xi_2\rangle=1$ | $\langle K_1\xi_2\xi_3\rangle=1$ |
$\langle K_1\xi_3\xi_4\rangle=1$|   \cr
$K_2$ | $\langle K_2\xi_1\xi_1\rangle=1$ | $\langle K_2\xi_1\xi_3\rangle=1$ |
$\langle K_2\xi_2\xi_2\rangle =\kappa^2$| $\langle K_2\xi_2\xi_3\rangle =\kappa$  \cr
$K_2$ | $\langle K_2\xi_3\xi_3\rangle=1$ | $\langle K_2\xi_3\xi_4\rangle=\kappa$ |
$\langle K_2\xi_4\xi_4\rangle =\kappa^2$|   \cr
$K_3/\widetilde{K}_3$ | $\langle K_3\xi_1\xi_5\rangle=1$ | $\langle K_3\xi_1\xi_6\rangle=1$ |
$\langle K_3\xi_2\xi_6\rangle =\kappa$| $\langle K_3\xi_3\xi_6\rangle =1$  \cr
| $\langle K_3\xi_3\xi_7\rangle=\kappa$ | ||\cr
$K_4/\widetilde{K}_4$ | $\langle K_4\xi_1\xi_7\rangle=1$ | $\langle K_4\xi_2\xi_5\rangle=1$ |
$\langle K_4\xi_2\xi_6\rangle =1$| $\langle K_4\xi_3\xi_7\rangle =1$  \cr
| $\langle K_4\xi_4\xi_6\rangle=1$ | ||\cr
$K_5/\widetilde{K}_5$ | $\langle K_5\xi_1\xi_6\rangle=1$ | $\langle K_5\xi_2\xi_6\rangle=\kappa$ |
$\langle K_5\xi_2\xi_7\rangle =\kappa^2$| $\langle K_5\xi_3\xi_5\rangle =1$  \cr
| $\langle K_5\xi_3\xi_6\rangle=1$ |$\langle K_5\xi_3\xi_7\rangle=\kappa$|$\langle K_5\xi_4\xi_6\rangle=\kappa$|$\langle K_5\xi_4\xi_7\rangle=\kappa^2$\cr
$K_6/\widetilde{K}_6$ | $\langle K_6\xi_3\xi_6\rangle=\kappa$ | $\langle K_6\xi_4\xi_7\rangle=\kappa^3$ ||
\endtable
\noindent
{\bf Table 1.}~{\it
Nonvanishing ring relations for the (80,0) B-model ground-ring. $K_n$ and $\widetilde{K}_n$ transform in the
{\bf 10} of SO(10) and $\xi_a$ and $\widetilde{\xi}_a$ in the {\bf 16}; $\kappa$ is a numerical constant. Where a $\widetilde{K}$ appears in the left-hand column, the $\widetilde{K}$ relation is the same as that for $K$ but with $\xi_5,\xi_6,\xi_7$ replaced with their $\widetilde{\xi}$ counterparts (adapted from \BlumenhagenEW).
}
\vskip12pt
The sufficiently curious reader
might find it entertaining to return to the literature on exactly solved (0,2)
models and identify a slew of topological ground rings; these might turn out
to be very handy in the study of (0,2) mirror symmetry.

\subsec{Mirror Symmetry and a Massive Model}

The worldsheet construction of mirror symmetry of (0,2) models
\AdamsZY\foot{See also \eg \BlumenhagenVU\BlumenhagenTV\SharpeWA\
for alternate approaches to (0,2) avatars of mirror symmetry.}
also provides strong evidence for the existence of these (0,2)
rings in both conformal and massive models - more precisely,
mirror symmetry has led to the prediction of such rings in
numerous systems. The most well-studied example involves the
deformation of the tangent bundle of $\IP^1 \times \IP^1$, a
massive (0,2) model whose A ring was computed via mirror symmetry
in \AdamsZY\ and checked via direct computation of the
intersection form on the associated instanton moduli space by Katz
and Sharpe\KatzNN\ (see also \SharpeFD).

The basic strategy of \AdamsZY\ involved the extension of Morrison-Plesser/Hori-Vafa\MorrisonYH\HoriKT\ worldsheet dualization techniques to construct dual pairs of (0,2) models related by the mirror automorphism, $\JL \leftrightarrow -\JL$ and $\QL \leftrightarrow \QLB$, with the mirror superpotential effectively summing the instantons of the original $\s$-model.  The ring derived from the mirror superpotential was thus interpreted as the mirror of a quantum cohomology of the original $\s$-model.

In the case of $T[\IP^1 \times \IP^1]$, the resulting ring relations turn out to be\foot{Using the notation of \KatzNN.}
$$
\tilde{X}^2 = \exp(it_2),\,\,\,\,\,
X^2-(\eps_1-\eps_2)X\tilde{X}=\exp(it_1)
$$
where $t_1,t_2$ are the K\"{a}hler parameters of the two $\IP^1$s,
while $\eps_1,\eps_2$ parametrize certain elements in $H^1(X,
End(T_X))$, the bundle moduli. Notice that the latter parameters
need not be perturbatively small.

Katz and Sharpe checked this argument by explicitly computing the cohomology products this ring was expected to reproduce \KatzNN, roughly generalizing the Gromov-Witten counting of rational curves to the (0,2) context and reducing the problem to precise computations in sheaf cohomology on the instanton moduli space.  Explicitly, since the anomaly ensures that each correlation function receives corrections from only specific worldsheet instanton sectors, by computing the appropriate intersection numbers on the instanton moduli space, Katz and Sharpe, in a remarkable and powerful computation, constructed by hand the two- and four-point functions\foot{Including, more recently, even the coefficients\SheldonWIP.} of the generators of the ground ring, finding precise agreement with the ring relations predicted in \AdamsZY.

With one important caveat - \KatzNN\ computed the intersection form on the moduli space of instantons with 2 or 4 marked points, but could not ensure that the associated A-model correlators were independent of the positions of the marked points.  Importantly, the results presented above imply that the OPE of A ring operators is non-singular, ensuring that these correlators are in fact independent of the insertion points.
The results of \KatzNN\ are thus in precise agreement with \AdamsZY.

\newsec{The Reservations at the End of the Paper}

In this paper we have verified the existence of topological ground-rings in A and B twisted (0,2) models in open balls around classical points in the moduli space, and globally under relatively generic assumptions, generalizing the quantum cohomology rings of (2,2) theories to the sheaf-cohomological context more natural to heterotic compactifications.  The key ingredient we needed was a left-moving $U(1)$ global symmetry by which to twist - in particular, we did not require the theory to be a deformation of a (2,2) model, nor even that it be geometrical: all arguments obtain at the level of the CFT, and thus cannot be destabilized by worldsheet instantons in the special case of non-linear $\s$-models.

It is quite remarkable that the topological rings of (2,2) theories persist as rings in (0,2) models under such mild constraints.  This leads to a host of natural questions.  What is the geometry of the malevolent operators which may destroy the ring when $\rk{\V}\geq8$ - are they related to conifolds, where the chiral ring also degenerates badly, or perhaps to some small heterotic instanton transition?  How are "bundle flops" realized in the chiral ring?  What are the generalizations of the periods, physically and mathematically?  In the case of massive models, can we make any global statements, or nail down the smooth behavior at infinity of A-model OPEs?  These and other questions we leave to future work.


\bigskip\noindent{\bf Acknowledgements}

\noindent
The authors thank
S.~Gukov,
D.~Morrison,
A.~Nietzke,
D.~Tong,
C.~Vafa,
S.~T.~Yau
and E.~Zaslow
and especially E. Sharpe and S. Katz for valuable discussions. Thanks also to S.~Bad for email correspondence.
A.~A. thanks the organizers and participants of the 2004 Fields Institute Workshop on Mirror Symmetry and the 2005 Amsterdam Summer Workshop where some of this work was discussed.  A.~A. and M.~E. thank the organizers and participants of the CMS String Workshop in Hangzhou.
J.~D. would like to thank the Perimeter Institute for hospitality while this work was being completed.
The work of A.~A. is supported by a Junior Fellowship from the Harvard Society of Fellows.
The work of J.~D. is supported by  the National Science Foundation under Grant Nos. PHY-0071512 and PHY-0455649, and by grant support from the US Navy, Office of Naval Research, Grant Nos. N00014-03-1-0639 and N00014-04-1-0336, Quantum Optics Initiative.

\bigskip
\bigskip
\bigskip
\bigskip
\bigskip
\vfill
\centerline{Dedicated to the memory of Rudolf Schrimpff.}

\listrefs

\end